\RequirePackage[OT1]{fontenc}
\documentclass[10pt,conference]{IEEEtran}
\usepackage{booktabs}
\usepackage{amsmath}
\usepackage{cite}
\usepackage{graphicx}
\usepackage{microtype}
\usepackage{xcolor}
\usepackage{url}

\graphicspath{{figures/}}

\newcommand{\system}{\textsc{DecompBench}}

\begin{document}

\bstctlcite{IEEEreferenceStyle}

\title{When Does Decomposition Help? Empirical Study on Workload--Architecture Affinity in LLM Inference Serving}

\author{\IEEEauthorblockN{Jihui Yang\textsuperscript{1},
Wentai Wu\textsuperscript{2},
Jiahua Huang\textsuperscript{3}*,
Weiwei Lin\textsuperscript{3},
Ligang He\textsuperscript{4},
Linghong Tsang\textsuperscript{2}
}
\IEEEauthorblockA{\textsuperscript{1}Meta Platforms, Inc., CA, USA\\
\textsuperscript{2}Jinan University, Guangzhou, China\\
\textsuperscript{3}South China University of Technology, Guangzhou, China\\
\textsuperscript{4}University of Warwick, Coventry, United Kingdom
}
}

\AddToHookNext{shipout/foreground}{%
  \put(0.675in,-10.18in){%
    \makebox(0,0)[lt]{\parbox[t]{7.15in}{%
      \normalfont\fontsize{8}{9}\selectfont\raggedright
      \textcopyright\ 2026 IEEE. Personal use of this material is permitted.
      Permission from IEEE must be obtained for all other uses, in any current
      or future media, including reprinting/republishing this material for
      advertising or promotional purposes, creating new collective works,
      for resale or redistribution to servers or lists, or reuse of any
      copyrighted component of this work in other works.}}}%
}

\maketitle

\begin{abstract}
Large language model (LLM) inference can be organized as independent
monolithic replicas, tensor- or pipeline-parallel serving groups, or
disaggregated prefill and decode services. Although decomposition is often
presented as a route to higher efficiency, its benefit depends on workload
shape, model fit, and whether phase-specific resources can be independently
scaled. Existing comparisons frequently change several of these factors at
once. We isolate the architectural effect by evaluating five SGLang serving
organizations under the same four-A100-40GB hardware envelope, pinned
Qwen3-8B model, software stack, prompts, and arrival schedules. Our evaluation
contains 150 audited conditions over short-chat, decode-heavy, and 32K
prefill-heavy workloads. Four independent one-GPU replicas achieve the lowest
end-to-end latency for short chat and long output and are the only architecture
stable in every condition. Pipeline parallelism reduces 32K time to first token
by 58--59\% at non-stress rates, while hybrid pipeline-based disaggregation
achieves the best 32K end-to-end latency at 0.4 requests/s. In contrast, B's
single-VM static TP2+TP2 split, without independent phase scaling, does
not improve tested latency over replication and collapses at higher 32K
load. These results
reveal a workload--architecture affinity: pipeline decomposition favors
prefill-heavy computation, whereas replication favors decode-heavy and
concurrent traffic. More broadly, decomposition is not a monotonic
optimization; its value depends on matching parallelism granularity to the
dominant phase and preserving queue stability.
\end{abstract}


\section{Introduction}

LLM inference consists of two computationally distinct phases. During
\emph{prefill}, the system processes all prompt tokens in parallel and is
typically compute intensive. During autoregressive \emph{decode}, it produces
one token at a time and is commonly limited by memory bandwidth. This
asymmetry has motivated systems that disaggregate the two phases, divide model
layers across devices, or use different parallelism strategies for each
phase~\cite{distserve,splitwise,orca,vllm}.

The case for decomposition is compelling but incomplete. Prefill/decode (PD)
disaggregation can reduce phase interference and enable independent scaling,
but it must transfer key-value (KV) state and can strand resources when a
fixed partition does not match demand. Pipeline parallelism (PP) can distribute
a large prefill computation while introducing bubbles during token-by-token
decode. Tensor parallelism (TP) reduces per-device model memory but adds
collective communication to every layer. Conversely, replication avoids these
costs when the model fits on one device, but is infeasible for larger models.

This interaction suggests a systems question that is easy to obscure in an
end-to-end deployment: \emph{when only the serving organization changes, which
workloads exhibit affinity toward which architecture?} We answer this question
inside a deliberately fixed envelope. Every condition uses the same four
A100-40GB GPUs, Qwen3-8B checkpoint, SGLang version, prompt sequence, arrival
timestamps, generation policy, and measurement client. We compare four
independent monolithic replicas, PD with TP2 per phase, unified PP4, hybrid
PD with PP2 per phase, and unified TP4.

Our study makes three contributions:

\begin{itemize}
  \item We present \system, a controlled methodology for comparing service
  decomposition while holding hardware, model, runtime, prompts, and arrivals
  fixed. The final matrix contains 150 conditions and passes strict prompt and
  arrival equality audits.
  \item We quantify workload--architecture affinity across short-chat,
  decode-heavy, and 32K prefill-heavy traffic. Replication is best for
  end-to-end decode performance, while PP provides the lowest long-context
  prefill latency.
  \item We identify two distinct failure modes under common offered load: a
  static TP2+TP2 PD split (B) leaves aggregate GPU capacity underused because one phase
  bottlenecks, whereas monolithic TP4 saturates all GPUs behind a growing
  scheduler queue.
\end{itemize}

The principal result is deliberately qualified. Our single-VM static
TP2+TP2 PD (B), without independent phase scaling, does not improve
tested P95 latency over replication. Pipeline C improves long-context
prefill, and hybrid PD (D) improves 32K E2E at 0.4 requests/s. This is
consistent with disaggregated systems that scale phase pools:
workload-aware provisioning and scaling freedom are essential.

We organize the evaluation around three research questions. \textbf{RQ1:} Does
the latency-optimal architecture change with workload shape? \textbf{RQ2:} Do
the advantages observed at low or moderate load survive a shared stress load?
\textbf{RQ3:} When an architecture fails, does it fail through full-device
saturation or through imbalance created by decomposition? These questions
separate phase-level speed from end-to-end service stability.

\section{Background and Motivation}

\subsection{The prefill--decode asymmetry}

For a prompt of length $L$ and a generated sequence of length $T$, prefill
exposes parallel work across the $L$ input tokens, whereas decode performs $T$
sequential model invocations. The optimal serving organization can therefore
depend on the ratio between input and output work. Long-context requests stress
prompt processing and KV-cache allocation; long-output requests stress
sustained generation; short chat stresses request-level concurrency and queue
management.

\subsection{Three forms of decomposition}

\textbf{Phase decomposition} places prefill and decode on separate workers and
transfers KV state between them. DistServe~\cite{distserve} and
Splitwise~\cite{splitwise} show the value of phase-specific placement and
scaling. \textbf{Layer decomposition} divides transformer layers across
pipeline stages, as introduced by GPipe~\cite{gpipe}. It changes both memory
placement and communication while potentially creating pipeline bubbles.
\textbf{Tensor decomposition} shards operations within each layer and exchanges
partial results through collectives, as formalized by Megatron-LM~\cite{megatron}.

These techniques solve different problems. Phase decomposition addresses
interference and scaling; PP and TP address model execution and memory. Their
combination need not be additive. We therefore treat architecture as an
experimental factor rather than assuming that finer decomposition is better.

\section{Architectures Under Test}

Table~\ref{tab:architectures} summarizes the five organizations. A and E form
the monolithic scale-out/scale-up controls. B isolates phase decomposition with
TP inside each phase. C isolates layer decomposition. D combines phase and
layer decomposition.

\begin{table}[!t]
\centering
\caption{Serving architectures under the same four-GPU envelope.}
\label{tab:architectures}
\small
\begin{tabular}{@{}cllc@{}}
\toprule
ID & Organization & Parallelism & GPUs \\
\midrule
A & Monolithic replicas & $4\times$TP1 & $1+1+1+1$ \\
B & Prefill/decode & TP2 + TP2 & $2+2$ \\
C & Layer-wise pipeline & PP4 & $4$ \\
D & PD + layer-wise & PP2 + PP2 & $2+2$ \\
E & Monolithic scale-up & TP4 & $4$ \\
\bottomrule
\end{tabular}
\end{table}

\begin{figure*}[t]
\centering
\includegraphics[width=\textwidth]{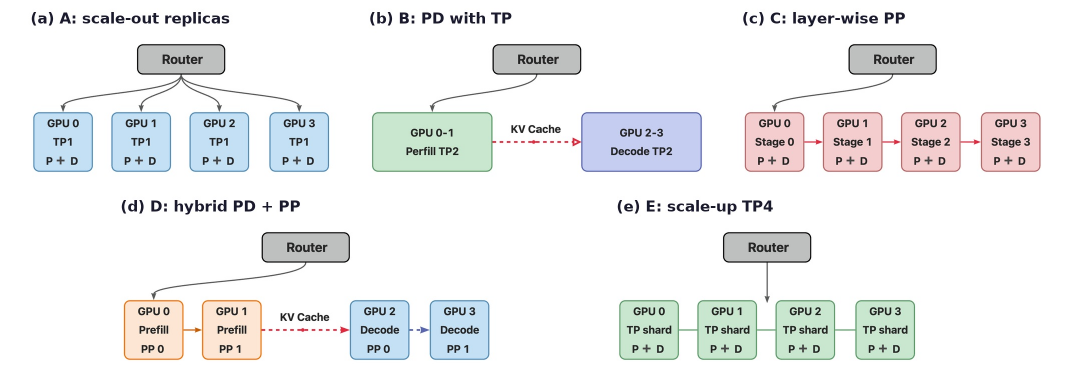}
\caption{Five serving organizations mapped to the same four GPUs. A and E are
monolithic scale-out/scale-up controls. B separates prefill and decode with TP
inside each phase. C uses one four-stage layer pipeline. D combines phase
separation with two-stage pipelines. Solid arrows denote request or activation
flow; dashed orange arrows denote KV transfer.}
\label{fig:architecture}
\end{figure*}

Architecture A exposes four independent model replicas behind one router.
Architecture E exposes one model replica synchronized across all four GPUs.
Because Qwen3-8B fits on one A100-40GB, this comparison directly measures the
cost of unnecessary scale-up. B and D run distinct prefill, decode, and router
processes in one VM. This controls external network variance but intentionally
removes independent multi-node autoscaling.

Figure~\ref{fig:architecture} makes explicit the synchronization boundaries.
In A, a request is routed to one GPU and remains there for both phases. In E,
every layer invokes TP collectives across all four devices. C also uses all
four GPUs for one request, but communicates stage activations rather than
partial tensor results. B and D introduce a second boundary: after prefill,
the KV state crosses from the prefill group to the decode group.

\subsection{Comparison logic}

The five cases form three controlled contrasts. A versus E asks whether an 8B
model that already fits one GPU should scale out through replicas or scale up
through TP. A versus B asks whether phase isolation compensates for halving the
GPU allocation available to each phase. C versus D asks whether adding phase
isolation to pipeline execution improves the balance between prompt processing
and generation. Finally, B versus D holds the 2+2 phase split fixed and changes
only the intra-phase parallelism from TP to PP.

This design does not rank all possible LLM deployments. Instead, it identifies
the local effect of decomposition inside a fixed allocation. In particular, B
and D cannot add prefill replicas when prompt demand rises. That restriction is
intentional: it lets us distinguish the overhead of decomposition from the
benefit of adding or independently scaling resources.

\section{Experimental Methodology}

\subsection{Controlled environment}

Every condition runs on one Vertex AI \texttt{a2-highgpu-4g} worker with four
A100-40GB GPUs. We use SGLang \texttt{v0.5.14-cu129} and pin Qwen3-8B~\cite{qwen3} to
revision \texttt{b968826d...91218}; the artifact records the full hash. The context limit
is 32,768 tokens. Radix cache remains enabled for every architecture. The
container digest, runtime packages, model revision, precision, router path,
generation settings, timeout, and telemetry collectors are fixed.

PD uses Mooncake's default transport with auxiliary TCP inside the VM. Thus,
the experiment evaluates scheduling, transfer, and static resource partitioning
without introducing cross-node network noise.

Table~\ref{tab:platform} records the frozen platform and software controls. We
disable CUDA graphs and SGLang custom all-reduce for every condition because
the common Vertex bring-up required that configuration; we also set
\texttt{NCCL\_P2P\_DISABLE=1} consistently. Enabling CUDA graphs and optimized collectives may change absolute
performance and relative rankings; this configuration remains untested.

\begin{table}[!t]
\centering
\caption{Frozen platform and serving configuration.}
\label{tab:platform}
\small
\begin{tabular}{@{}ll@{}}
\toprule
Control & Value \\
\midrule
Accelerators & $4\times$ A100-40GB \\
Host & 48 vCPUs, 340 GiB RAM \\
Model & Qwen3-8B, dense, BF16 \\
Serving runtime & SGLang \texttt{v0.5.14-cu129} \\
PD transport & Mooncake default + aux TCP \\
\bottomrule
\end{tabular}
\end{table}

\begin{table*}[t]
\centering
\caption{Workloads, offered rates, and SLOs. Every pool contains 5,000 unique
prompts; output lengths are maximum generation caps.}
\label{tab:workloads}
\small
\begin{tabular*}{\textwidth}{@{\extracolsep{\fill}}llrrlrr@{}}
\toprule
Workload & Dominant pressure & Input tokens & Output cap & Rates (req/s) & TTFT SLO & E2E SLO \\
\midrule
Short chat & concurrency/queueing & 9--2,034 & 64 & 4, 8, 12 & 2 s & 10 s \\
Long output & sustained decode & 10--4,026 & 512 & 0.5, 1, 2 & 3 s & 60 s \\
32K context & prefill/KV allocation & 31,990--31,995 & 32 & 0.05, 0.1, 0.4 & 15 s & 60 s \\
32K stress & prefill saturation & 31,990--31,995 & 32 & 0.8 & 15 s & 60 s \\
\bottomrule
\end{tabular*}
\end{table*}

\begin{table}[!t]
\centering
\caption{Replay, measurement, and statistical protocol.}
\label{tab:protocol}
\small
\begin{tabular}{@{}ll@{}}
\toprule
Parameter & Frozen value \\
\midrule
Arrival interval & 300 s \\
Warmup / stable window & 120 s / 180 s \\
Client workers & 256 \\
Per-request timeout & 180 s \\
Replicates & 3 deterministic seeds \\
GPU sampling & 2 s \\
Server-metric sampling & 10 s \\
Scheduled requests & 129,825 total \\
Accepted result records & 150/150 \\
Uncertainty & 95\% Student-$t$ CI \\
Cost basis & common stable window \\
\bottomrule
\end{tabular}
\end{table}

\subsection{Workloads and shared rates}

We construct three committed pools of 5,000 normalized, deduplicated prompts.
They represent conversational traffic informed by LMSYS-Chat-1M~\cite{lmsys}, decode-heavy generation, and
prefill-heavy 32K context. The content is derived from conversational and
long-form public corpora; the benchmark evaluates serving, not answer quality.
Long-context examples are assembled and tokenized to a narrow 31,990--31,995
token band so that architecture, rather than prompt length, explains prefill
variation. Table~\ref{tab:workloads} lists the token ranges, shared offered
rates, and frozen SLOs.

Each condition schedules arrivals for 300 seconds. Three deterministic seeds
select prompts without replacement. For every matching seed, workload, and
rate, all architectures receive exactly the same ordered request IDs and
arrival offsets. This shared-rate design answers the operator-facing question:
given the same traffic, which architecture serves it best?

Large prompt pools are important because SGLang's radix cache remains enabled.
Early synthetic pilots reused a small number of templates and consequently
overstated cache benefits. In the accepted matrix, prompts are unique within a
condition, sampled without replacement, and shared exactly across A--E. Natural
prefix overlap is therefore preserved as a realistic runtime feature while its
effect remains controlled across architectures.

\begin{figure*}[t]
\centering
\includegraphics[width=\textwidth]{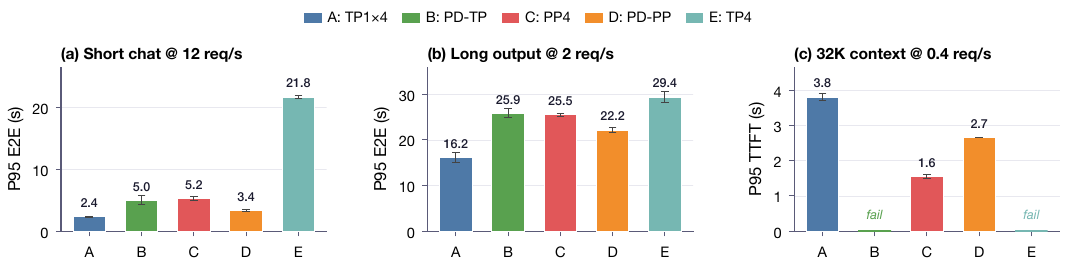}
\caption{Workload--architecture affinity at representative shared rates.
(a--b) Independent replicas minimize E2E latency for short-chat and
decode-heavy traffic. (c) PP4 minimizes 32K TTFT; B and E fail the condition.
Bars show three-seed means and 95\% confidence intervals.}
\label{fig:affinity}
\end{figure*}

\subsection{Metrics and pass criterion}

Under the protocol in Table~\ref{tab:protocol}, the first 120 seconds form a warmup; the following 180 seconds form the stable
window. A TP4 short-chat pilot with 60-second warmup showed cold-start
backlog through 120 seconds. We extended warmup while retaining
180 seconds for measurement (six trend buckets) before freezing the
matrix. This excludes observed transients without establishing long-run
stationarity. The client records request-level time to first token (TTFT), end-to-end
(E2E) latency, token counts, and errors. GPU telemetry captures utilization,
memory, and power every two seconds; SGLang metrics capture queue depth, running
requests, cache hit rate, and generation throughput every ten seconds.

A condition passes only when (1) error rate is at most 1\%, (2) the endpoint
and all serving processes remain healthy, (3) stable-window P95 TTFT and E2E
meet the workload SLO, and (4) arrival delay and TTFT show no sustained upward
trend. The TTFT/E2E SLOs are 2/10 seconds for short chat, 3/60 seconds for long
output, and 15/60 seconds for 32K context. Completion and stability are
separate: a condition may complete all requests yet fail because its queue is
growing.

The final audit validates all 150 condition artifacts, immutable identities,
request rows, and telemetry. We report three-seed means; the artifact tables
also include 95\% Student-$t$ confidence intervals.

\subsection{Derived metrics and stability}

For condition $c$, let $S_T$ and $S_E$ denote the TTFT and E2E SLO thresholds,
respectively. Let $\mathcal{R}_W^{\rm ok}$ contain successful requests scheduled
in the stable window $W$, including completions after $W$. Stable goodput counts
those meeting both SLOs:
\begin{equation}
G_c = \frac{1}{|W|}\sum_{r\in\mathcal{R}_W^{\rm ok}}
\mathbf{1}[\mathrm{TTFT}_r\leq S_T \wedge \mathrm{E2E}_r\leq S_E].
\end{equation}
Nominal cost per 1K output tokens divides the fixed worker cost over $W$ by all
successful completion tokens from that arrival cohort, including tokens from
requests that miss an SLO. Cost per SLO-good request instead divides the same
cost by the number of requests meeting both SLOs.

Our pass gate is intentionally stricter than an HTTP-success gate. We divide
the stable window into 30-second buckets and reject sustained growth in arrival
delay or P95 TTFT. This distinguishes a temporarily completed trace from a
service point that can remain stationary. The endpoint must also remain healthy,
all processes must be live, and error rate must not exceed 1\%.

\section{Results}

\subsection{Stability across workloads}

Table~\ref{tab:passes} reports passing replicates out of three. A passes all 30
replicates across its 10 conditions. Long output does not separate capacity within the tested
range, but short chat exposes E's scale-up bottleneck and long context sharply
separates all five architectures.

\begin{table}[!ht]
\centering
\caption{Passing replicates out of three.}
\label{tab:passes}
\small
\setlength{\tabcolsep}{3pt}
\begin{tabular}{@{}lcccccccccc@{}}
\toprule
& \multicolumn{3}{c}{Short chat} & \multicolumn{3}{c}{Long output} &
\multicolumn{4}{c}{32K context} \\
\cmidrule(lr){2-4}\cmidrule(lr){5-7}\cmidrule(l){8-11}
Architecture & 4 & 8 & 12 & 0.5 & 1 & 2 & 0.05 & 0.1 & 0.4 & 0.8 \\
\midrule
A: TP1$\times$4 & 3 & 3 & 3 & 3 & 3 & 3 & 3 & 3 & 3 & 3 \\
B: PD TP2+TP2 & 3 & 3 & 3 & 3 & 3 & 3 & 3 & 3 & 0 & 0 \\
C: PP4 & 3 & 3 & 3 & 3 & 3 & 3 & 3 & 3 & 3 & 2 \\
D: PD PP2+PP2 & 3 & 3 & 3 & 3 & 3 & 3 & 3 & 3 & 3 & 0 \\
E: TP4 & 3 & 1 & 0 & 3 & 3 & 3 & 3 & 3 & 0 & 0 \\
\bottomrule
\end{tabular}
\end{table}

Across the full matrix, 129 of 150 conditions pass. All 45 long-output
conditions pass, 40 of 45 short-chat conditions pass, and 44 of 60 long-context
conditions pass. Output-heavy traffic therefore separates latency efficiency,
whereas 32K traffic separates both latency and capacity.

\begin{table*}[t]
\centering
\caption{Representative three-seed mean results. ``Fail'' denotes no valid
stable-window latency because the condition timed out or violated the pass
gate. Best values among passing architectures are bold.}
\label{tab:keyresults}
\small
\begin{tabular*}{\textwidth}{@{\extracolsep{\fill}}lrrrrrrr@{}}
\toprule
& \multicolumn{2}{c}{Short 12 req/s} & \multicolumn{2}{c}{Output 2 req/s} &
\multicolumn{2}{c}{32K 0.4 req/s} & 32K 0.8 req/s \\
\cmidrule(lr){2-3}\cmidrule(lr){4-5}\cmidrule(lr){6-7}\cmidrule(l){8-8}
Architecture & TTFT & E2E & TTFT & E2E & TTFT & E2E & Error rate \\
\midrule
A: TP1$\times$4 & 0.136 & \textbf{2.375} & 0.108 & \textbf{16.195} & 3.806 & 4.702 & \textbf{0.0\%} \\
B: PD TP2+TP2 & 1.490 & 5.043 & 0.338 & 25.890 & Fail & Fail & 92.9\% \\
C: PP4 & \textbf{0.135} & 5.249 & \textbf{0.087} & 25.533 & \textbf{1.554} & 4.240 & \textbf{0.0\%} \\
D: PD PP2+PP2 & 0.369 & 3.358 & 0.331 & 22.176 & 2.655 & \textbf{3.982} & 6.9\% \\
E: TP4 & 0.974 & 21.763 & 0.222 & 29.407 & Fail & Fail & 89.6\% \\
\bottomrule
\end{tabular*}
\end{table*}

\begin{figure*}[t]
\centering
\includegraphics[width=\textwidth]{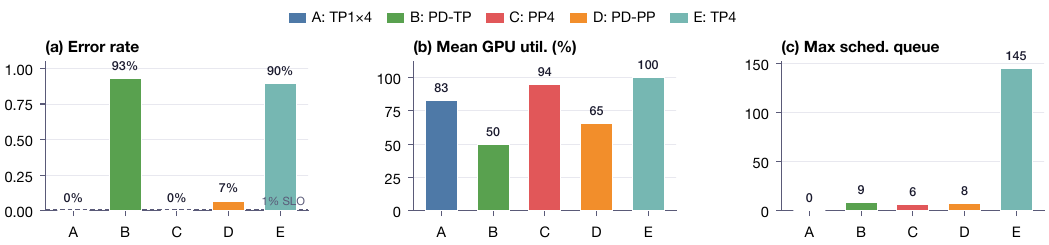}
\caption{Failure anatomy at the common 32K/0.8 requests/s stress load.
B fails with only 50\% mean GPU utilization, indicating static phase imbalance;
E saturates the GPUs behind a queue of 145. A remains stable.}
\label{fig:stress}
\end{figure*}

\subsection{Replication favors short and decode-heavy traffic}

Figure~\ref{fig:affinity}a compares latency at 12 short-chat requests/s. A reaches
2.375 seconds P95 E2E, compared with 3.358 for D, 5.043 for B, 5.249 for C,
and 21.763 for E. Relative to E, A reduces P95 E2E by 61.7\%, 78.1\%, and
89.1\% at 4, 8, and 12 requests/s. E completes the 12 requests/s trace but
violates the 10-second E2E SLO, illustrating why throughput alone is
insufficient.

The same ordering appears for long output. At 2 requests/s, A's P95 E2E is
16.195 seconds, followed by D at 22.176, C at 25.533, B at 25.890, and E at
29.407. C provides the lowest TTFT (0.087 seconds), but its pipeline decode
path extends generation time. This distinction is central: the architecture
that starts fastest need not finish fastest.

The load response reinforces this result. A's short-chat P95 E2E changes only
from 2.110 to 2.375 seconds as offered load triples from 4 to 12 requests/s. E
increases from 5.502 to 21.763 seconds over the same range. For long output,
A remains nearly flat at 15.769--16.195 seconds, while E rises from 25.433 to
29.407 seconds. Independent replicas provide both a lower latency intercept and
a smaller queueing slope when decode occupies a substantial service fraction.

\subsection{Pipeline decomposition favors long prefill}

At 0.05, 0.1, and 0.4 32K requests/s, C's P95 TTFT is 1.591, 1.566, and 1.554
seconds respectively, 58--59\% lower than A at each rate. At 0.4 requests/s (Table~\ref{tab:keyresults}), D provides the best P95 E2E at
3.982 seconds, compared with 4.240 for C and 4.702 for A. Pipeline/layer
decomposition therefore exhibits affinity toward compute-intensive prefill,
and the hybrid architecture balances this gain with decode isolation at
moderate load.

B does not share this advantage. Even at 0.05 requests/s, its 8.474-second
P95 TTFT is 2.23$\times$ A's. At 0.4 requests/s, each B replicate completes
exactly 21 of the 120 total scheduled requests; E completes exactly 25. These
identical plateaus across prompt seeds indicate deterministic service-capacity
limits rather than cloud noise.

At the lowest 32K rate, C also has the best E2E latency (2.846 seconds),
followed by D (3.804), A (4.687), E (8.611), and B (9.995). The ordering shows
that PP's prefill advantage is not merely a near-saturation effect. However,
C's gap over A narrows as load rises: C reaches 37.122-second E2E at 0.8
requests/s, whereas A remains close to 4.7 seconds throughout. Affinity must
therefore be stated jointly with the operating region.

\subsection{Stress reveals different failure modes}

At the common 0.8 32K requests/s stress rate (Figure~\ref{fig:stress}), A is the only architecture that
passes all replicates. It maintains 3.823-second P95 TTFT, 4.732-second P95
E2E, and zero queue buildup. C completes 240/240 requests in each seed
with zero errors and meets the absolute 15/60-second SLOs; its mean P95
E2E is 37.122 seconds (Table~\ref{tab:seeds}). Seed 20260608 alone fails
the TTFT trend gate: its 1.649 seconds/minute slope and 4.121-second
first-to-last increase both exceed the frozen thresholds of
1.5 seconds/minute and 3.0 seconds, respectively.

D reaches 175.696-second P95 E2E with 6.94\% errors. B and E fail 92.92\% and
89.58\% of requests. Yet their telemetry differs: B averages only 50.1\% GPU
utilization while failing, consistent with one overloaded phase and stranded
capacity in the other. E averages 99.7\% utilization and reaches a normalized
maximum scheduler queue of 145, consistent with saturation of one TP4 replica.

\begin{table}[!ht]
\centering
\caption{C (PP4) at 32K/0.8 requests/s: per-seed P95
latencies and TTFT trends.}
\label{tab:seeds}
\small
\setlength{\tabcolsep}{3pt}
\begin{tabular}{@{}rrrrrc@{}}
\toprule
Seed & TTFT & E2E & Slope & Increase & Pass \\
 & (s) & (s) & (s/min) & (s) & \\
\midrule
20260606 & 9.560 & 37.583 & 0.916 & 2.204 & Yes \\
20260607 & 8.454 & 36.495 & 0.707 & 1.744 & Yes \\
20260608 & 10.114 & 37.288 & 1.649 & 4.121 & No \\
\bottomrule
\end{tabular}
\end{table}

\subsection{Cost and SLO goodput}

The formal matrix records 18.283 worker-hours, or \$293.54 under the frozen
price model. Since every condition uses the same four-GPU worker, stable runs
at a common rate have nearly identical nominal cost per output token. At
32K/0.4, A, C, and D each cost \$0.3484 per 1K successful output tokens from the
stable-window arrival cohort; all those requests also meet both SLOs. B and E
produce no successful output in that cohort.

At lower rates, successful output-token volumes and nominal unit costs are
nearly equal. An architecture can retain successful output while losing
SLO-goodput. Cost per SLO-good request is undefined when no requests meet both
SLOs, even if nominal token cost remains finite. These costs reflect service
capacity under a fixed worker price rather than device energy.

\section{Understanding Workload--Architecture Affinity}

The results support a two-dimensional interpretation. First, the ratio of
prefill to decode work determines which organization minimizes latency.
Layer-wise PP accelerates a large prefill, whereas independent replicas provide
more concurrent decode streams and avoid per-token synchronization. Second,
offered load determines whether this phase-level advantage survives queueing.
C has the best 32K TTFT at moderate rates, yet A is more robust at stress.

\textbf{Why static TP2+TP2 PD (B) does not win.} The central systems benefit of PD is
independent phase scaling. B removes this degree of freedom by assigning
exactly two GPUs to each phase. KV transfer and TP synchronization add latency,
while the partition cannot redirect an idle decode GPU to an overloaded
prefill phase. D improves long prefill through PP but reaches the same static
partition boundary under stress.

\textbf{Why replication is strong here.} Qwen3-8B fits on one A100-40GB. A can
therefore turn four GPUs into four independent execution streams. This is a
model-fit-dependent result, not a universal recommendation. If model weights
or KV state require multiple GPUs, A becomes infeasible and the feasible
architecture set changes.

\textbf{Operational implication.} For an 8B model on four A100-40GB GPUs,
replication is a strong default for mixed or decode-heavy traffic. PP is
attractive when long-context TTFT dominates. Fixed equal PD splits should be
treated as capacity-planning decisions, not automatic optimizations; they need
phase-specific demand estimates or elastic replica pools.

\subsection{A phase-capacity interpretation}

Let $\lambda$ be offered request rate and let $\mu_P$ and $\mu_D$ denote the
service capacities of the prefill and decode pools. A static PD system is stable
only when $\lambda < \min(\mu_P,\mu_D)$. Allocating two GPUs to each phase is
optimal only when that allocation equalizes the two capacities for the target
workload. The 32K workload increases prefill work without increasing the
32-token decode cap, pushing $\mu_P$ below $\mu_D$. B's low aggregate GPU
utilization during failure is the observable consequence: the decode half
cannot consume its reserved capacity because the prefill half cannot supply KV
states quickly enough.

The monolithic cases have no fixed phase boundary. In A, the router distributes
whole requests across four servers; an idle GPU can begin the next request
regardless of its phase. E also avoids a static phase split, but one TP4 replica
serializes request progress through collective operations. Its high GPU
utilization and queue growth show a different bottleneck: all devices are busy,
yet the system exposes insufficient independent request streams.

\subsection{TTFT and E2E can disagree}

C minimizes TTFT for long output and 32K context at the representative high
rates (the short-chat TTFT difference between A and C is within 0.001~s),
but A minimizes E2E for short chat and long output. PP distributes the prompt's layer
work across stages and can begin output quickly, yet each generated token must
continue through the same stage sequence. For long output, this repeated
pipeline traversal dominates the one-time prefill gain. D reduces E2E relative
to C by isolating decode, but its two-stage decode pipeline remains slower than
four independent TP1 decoders.

This distinction matters operationally. A chat product optimizing first-token
responsiveness may choose differently from a batch-generation service
optimizing completion latency, even under the same request-rate SLO. Reporting
only TTFT or only throughput would hide that trade-off.

\subsection{Decision boundaries}

Within the tested scope, an operator can use a simple decision sequence. If the
model fits one GPU and completion latency or mixed traffic dominates, begin with
replication. If prompt processing dominates and load remains below the measured
pipeline saturation point, PP can materially improve TTFT. Use PD when its
primary operational benefit is available---independent phase replica counts,
phase-aware routing, or heterogeneous placement---and size those pools from
measured phase demand. A symmetric split is a hypothesis to test, not a neutral
default. Table~\ref{tab:selection} summarizes these guidelines.

\begin{table}[t]
\centering
\caption{Architecture selection guide within the tested scope.}
\label{tab:selection}
\small
\setlength{\tabcolsep}{4pt}
\begin{tabular}{@{}p{2.6cm}ll@{}}
\toprule
Workload profile & Recommended & Rationale \\
\midrule
Short chat / mixed & A (TP1$\times$4) & Lowest E2E; best stability \\
Long output & A (TP1$\times$4) & Lowest E2E; flat load curve \\
Long context (mod.) & C (PP4) & 58\% lower TTFT \\
Long context (high) & A (TP1$\times$4) & Only arch.\ stable at stress \\
TTFT \& E2E & D (PD-PP) & Best 32K E2E at 0.4 req/s \\
Elastic phase pools & B/D + scaling & Requires pool sizing \\
\bottomrule
\end{tabular}
\end{table}

\subsection{Implications for adaptive serving}

The matrix suggests that a practical deployment should expose architecture as
a routing and provisioning decision rather than a cluster-wide constant. A
front end can classify requests using input length, requested output cap, and
current phase queues. Short and decode-heavy requests can be sent to replicated
monolithic workers, while very long prompts can be directed to a PP or PD pool
when its measured TTFT benefit exceeds transfer and queueing overhead. This is
not per-request ``oracle'' selection: the three fields are available before
generation, and queue depth is already maintained by serving runtimes.

Elastic PD should further size prefill and decode pools from phase demand. If
$w_P$ and $w_D$ are measured GPU-seconds per request for each phase, a first
allocation target is $n_P/n_D \approx w_P/w_D$, followed by an SLO-constrained
search that accounts for KV-transfer cost and batching. Our results show why a
fixed $n_P=n_D$ policy is fragile: 32K prompts change $w_P$ by orders of
magnitude without a corresponding increase in decode work. Production systems
such as DistServe, Splitwise, and Mooncake use phase-aware placement, scheduling,
or distributed KV storage precisely to recover this missing degree of
freedom~\cite{distserve,splitwise,mooncake}.

If TP2 is the smallest feasible replica, four GPUs provide two monolithic
replicas. Static TP2+TP2 PD has one replica per phase and remains limited
by the slower phase. Larger-model execution, batching, and communication
can change these capacities and reverse the TP1-based ranking; this
regime was not evaluated here.

\section{Reproducibility and Artifact Integrity}

The artifact bundle contains 150 raw JSON records, request-level rows, GPU
samples, and SGLang Prometheus samples. Every worker records the immutable
container digest, code bundle hash, model revision, and runtime versions. The
strict auditor rejects missing keys, prompt or arrival mismatches, hardware
drift, unhealthy endpoints, and hidden streaming aborts. The audit proves exact
request-ID and arrival-order equality across A--E for every matching seed,
workload, and rate. The formal matrix schedules 129,825 requests and records
127,879 successful responses. One normalization is applied: PP ranks export
duplicate scheduler gauges whose sum we divide by the rank count before
comparing queue depth; all client-side metrics are independent of this step.

\section{Related Work}

\subsection{Disaggregated LLM serving}

Recent serving systems disaggregate prefill and decode to optimize latency and hardware utilization~\cite{hu2025shuffleinfer,liao2026dopd}. DistServe~\cite{distserve} disaggregates prefill and decode and jointly searches
parallelism and placement under latency SLOs. Splitwise~\cite{splitwise}
similarly exploits phase asymmetry and heterogeneous hardware. Mooncake adds a
distributed KV-cache hierarchy and KV-aware global scheduling for
long-context production traffic~\cite{mooncake}. Our fixed
four-GPU study evaluates the complementary boundary in which independent pool
scaling is unavailable, exposing the local overhead and imbalance of phase
separation.

AMPD~\cite{AMPD} coordinates prefill workloads adaptively across disaggregated
workers for multi-round inference. Infinite-LLM~\cite{Infinite-LLM}
disaggregates attention layers and pools KV cache across a cluster to support
dynamic long-context serving. A recent Meta study~\cite{Buzz} evaluates
disaggregation pragmatically, finding chunked pipeline parallelism and
Pareto-optimal batch sizing key to balancing throughput and latency. Prior PD
systems seek an optimized deployment and may change replica counts to satisfy
an SLO. We freeze the total allocation and replay identical traffic to measure
what phase separation itself contributes; our negative fixed-split result helps
explain why workload-aware provisioning is necessary.

\subsection{Serving runtimes and parallelism}

Orca~\cite{orca}, vLLM~\cite{vllm}, and SGLang~\cite{sglang} improve serving
through iteration-level scheduling, paged KV memory, prefix reuse, and
structured execution. DeepSpeed-FastGen~\cite{FastGen} composes prompt
processing and token generation into unified batches via Dynamic SplitFuse.
FlashAttention~\cite{FlashAttention} and FlashInfer~\cite{FlashInfer}
provide IO-aware and customizable attention kernels that underlie modern
serving runtimes. We hold the serving runtime fixed to SGLang so these
engine-level choices do not confound architecture. Prior benchmarking work such
as MLPerf Inference~\cite{mlperf} emphasizes reproducible system comparison;
our contribution is an exact cross-architecture replay and stability audit for
decomposed LLM serving.

Model-parallel runtimes commonly select TP, PP, or combinations to fit a model
or increase throughput~\cite{megatron2021}. Our 8B setting removes model fit as
a necessity, turning C and E into controls for PP and TP communication effects.
AlpaServe~\cite{alpaserve} studies statistical multiplexing across models;
Sarathi-Serve~\cite{sarathi} reduces pipeline bubbles through chunked prefill;
Llumnix~\cite{Llumnix} adds dynamic request migration for load balancing;
FlexGen~\cite{flexgen} enables high-throughput offline inference through memory
offloading. We leave scheduler changes disabled so the measured PP effect
comes from organization rather than a specialized pipeline scheduler.

\subsection{Benchmark methodology}

LLM serving studies typically report latency-throughput curves, token
throughput, or SLO goodput. Our methodology adds two controls for architecture
affinity: exact prompt/arrival equality and an explicit queue-trend gate. The
former prevents prompt length and prefix-cache variation from changing with
architecture; the latter prevents a finite trace from labeling a
non-stationary operating point as sustainable.

\section{Limitations and Threats to Validity}

\textbf{Internal validity.} Cloud variability is mitigated through three prompt
seeds, exact replay, immutable environment checks, and dual client/server
signal collection. The repeated completion plateaus for B and E and their
distinct GPU/queue signatures make capacity failures unlikely to be transient
noise. Three replicates provide limited power for small differences; we
therefore emphasize large effects and include confidence intervals.

\textbf{Construct validity.} TTFT and E2E do not capture output quality,
fairness, or recovery after overload. Our frozen SLOs are application choices;
``best'' should be read relative to the stated workload and SLO.

\textbf{External validity.} Results cover one dense 8B model on four
A100-40GB GPUs under one SGLang release. Larger models, quantization,
H100-class devices, multi-node RDMA, or elastic phase pools can reverse the
trade-offs. PD processes share one VM, so independent autoscaling and cluster
scheduling are not evaluated. Each trace lasts five minutes; longer production
traces may expose slower dynamics.

\section{Conclusion}

Service decomposition is not a monotonic optimization. In the fixed
four-A100 Qwen3-8B setting, independent replicas minimize short-chat and
long-output E2E latency and provide the strongest queue stability. PP
improves 32K prefill latency; hybrid PD improves moderate-load 32K E2E,
but both incur decode or saturation costs. Static TP2+TP2 PD (B) adds
overhead and phase imbalance without independent scaling. Select
decomposition by workload, model fit, and scaling freedom, and validate
it under shared offered load rather than phase-level latency alone.

\section*{Acknowledgments}
This work is supported by New Generation Artificial Intelligence National Science and Technology Major Project (2025ZD0123605), National Natural Science Foundation of China (62402198), Basic and Applied Basic Research Foundation of Guangzhou (2025A04J2212), Guangdong Provincial Natural Science Foundation Project (2025A1515010113), Shandong Provincial Natural Science Foundation Project (ZR2024LZH012), Guangxi Key Research and Development Project (2024AB02018), and the Major Key Project of PCL, China under Grant PCL2025A11.

\bibliographystyle{IEEEtran}
\IEEEtriggeratref{8}
\bibliography{references,ieee-reference-style}

\end{document}